\documentclass[11pt,a4paper]{article}

\usepackage{amsmath}
\usepackage{amssymb}
\usepackage{mathrsfs}
\usepackage{cite}
\usepackage{braket}
\usepackage[top=25mm,bottom=25mm,left=25mm,right=25mm]{geometry}
\usepackage[colorlinks=true,linkcolor=blue,urlcolor=blue,citecolor=blue]{hyperref}

\DeclareMathOperator{\Tr}{Tr}  

\begin{document}
\thispagestyle{empty}
\begin{flushright}
\tt CP3-16-43
\end{flushright}
\vskip 10pt

\begin{center}
{\huge\sffamily\bfseries A note on connected formula for form factors}
\vskip 30pt

{\large\sc Song He}
\\
\vskip 1pt
{\it CAS Key Laboratory of Theoretical Physics,
Institute of Theoretical Physics, \\
Chinese Academy of Sciences,
Beijing 100190, P. R. China}
\\
\vskip 1pt
{\it Email:\,}{\tt\color{blue} songhe@itp.ac.cn}

\vskip 15pt
{\large\sc Zhengwen Liu}
\\
\vskip 1pt
{\it Center for Cosmology, Particle Physics and Phenomenology (CP3), \\
Universit\'{e} catholique de Louvain, B1348 Louvain-la-Neuve, Belgium}
\\
\vskip 1pt
{\it Email:\,}{\tt\color{blue} zhengwen.liu@uclouvain.be}
\end{center}

\vskip 15mm
{\noindent\large {\bf Abstract:}~~In this note we study the connected prescription, originally derived from Witten's twistor string theory, for tree-level form factors in ${\cal N}=4$ super-Yang-Mills theory. The construction is based on the recently proposed four-dimensional scattering equations with $n$ massless on-shell states and one off-shell state, which we expect to work for form factors of general operators. To illustrate the universality of the prescription, we propose compact formulas for super form factors with chiral stress-tensor multiplet operator, and bosonic ones with scalar operators ${\rm Tr}(\phi^m)$ for arbitrary $m$.

\vskip 80mm
\noindent{\sc\today}
}
\newpage

\section{Introduction}

Witten's twistor string theory has provided a world-sheet model for the complete tree-level S-matrix in ${\cal N}\!=\!4$ supersymmetric Yang-Mills (SYM)~\cite{Witten:2003nn}. By performing half-Fourier transform from twistor space back to momentum space, the connected prescription of Roiban-Spradlin-Volovich (RSV)~\cite{Roiban:2004yf} expresses any $n$-point SYM amplitude as a localized integral over the moduli space of $n$-punctured Riemann spheres.
Recent progress along this line has been partly motivated by the scattering-equation-based construction~\cite{Cachazo:2013gna, Cachazo:2013hca} for S-matrices in any dimensions for various massless QFT's~\cite{Cachazo:2013iea, Cachazo:2014nsa, Cachazo:2014xea}.
The construction has been generalized to loop level~\cite{Geyer:2015bja, Geyer:2015jch, Cachazo:2015aol, Geyer:2016wjx} and derived from the so-called ambitwistor string theory~\cite{Mason:2013sva}. When reduced to four dimensions, one obtains old and new connected formulas, such as those for ${\cal N}\!=\!4$ SYM and ${\cal N}\!=\!8$ supergravity~\cite{Roiban:2004yf, Cachazo:2012da, Cachazo:2012kg,Cachazo:2012pz, Skinner:2013xp, Cachazo:2013iaa, Geyer:2014fka}, as well as for a large variety of theories including DBI-Volkov-Akulov theory~\cite{Adamo:2015gia, Cachazo:2016sdc, He:2016vfi, Cachazo:2016njl}.

What is special in four dimensions is that the scattering equations naturally split into $n-3$ sectors~\cite{Cachazo:2013iaa}, labeled by $k=2, 3,\ldots, n-2$, and for each solution sector they reduce to the equations of the original connected prescription in the corresponding helicity sector $k$ (for superamplitude this is the sector with Grassmann degree $(4 k)$). It is convenient to use a specific form of four-dimensional equations equivalent to RSV-Witten equations~\cite{Geyer:2014fka, He:2016vfi}: for sector $k$ we choose to split the $n$ particles into a set of $k$ particles we call ``$-$'', and the complimentary set ``$+$'' with $n{-}k$ particles. We have $2n$ equations for $2n$ variables $\sigma_a^{\alpha=1,2}$ for $a=1,2,\ldots,n$:
\begin{align}\label{eq}
  \tilde\lambda^{\dot\alpha}_I \,-\, \sum_{i\,\in\,+} \frac{\tilde\lambda^{\dot\alpha}_i}{(I\,i)}\,=\,0\,,\quad I\in -\,;
  \qquad
  \lambda^{\alpha}_i\,-\,\sum_{I\,\in\,-} \frac{\lambda^{\alpha}_I}{(i\,I)}\,=\,0\,,\quad i\in +\,.
\end{align}
Here and throughout the paper we use index $I,i$ for labels in the two sets $-,+$, and denote $(a\,b):=\epsilon_{\alpha\beta}\,\sigma_a^{\alpha} \sigma_b^{\beta}$. A connected formula expresses tree amplitudes as integrals over $\sigma$'s which are localized on the support of solutions of \eqref{eq}; for Yang-Mills amplitudes with $k$ negative helicities in $-$, we have the simplest formula whose integrand is just a  ``Parke-Taylor" factor:
\begin{align}\label{YM}
{\cal A}_{n,k}=\int \frac{\prod_{a=1}^n\,d^2\sigma_a}{{\rm vol\,GL}(2)~(12)(23)\cdots (n1)}~\prod_{I\,\in\,-} \delta^2\left(\tilde\lambda_I-\sum_{i\,\in\,+} \frac{\tilde\lambda_i}{(I\,i)}\right) \prod_{i\,\in\,+} \delta^2\left(\lambda_i-\sum_{I\,\in\,-} \frac{\lambda_I}{(i\,I)}\right)\,,
\end{align}
where we mod out the GL$(2,\mathbb{C})$ redundancy: four of the variables are fixed, and four redundant delta functions can be pulled out to impose the overall momentum conservation. Very recently, the connected formula, \eqref{YM}, has been extended to amplitudes in the standard model, such as massless quark-gluon amplitudes and amplitudes with a Higgs and multiple partons~\cite{He:2016dol}. The result for Higgs plus multi-gluon amplitude is particularly interesting for our purposes, because it opens up the exciting possibility for applying connected prescription to off-shell quantities.

Form factors provide a bridge between on-shell amplitudes and purely off-shell correlation functions, thus they are perfect for testing the applicability of on-shell techniques to off-shell generalizations. Recent years have witnessed considerable advances in the study of form factors in ${\cal N}\!=\!4$ SYM using on-shell methods and integrability, both at weak coupling \cite{
Brandhuber:2010ad,Bork:2010wf,
Brandhuber:2011tv,
Bork:2011cj,
Henn:2011by,
Gehrmann:2011xn,
Brandhuber:2012vm,
Bork:2012tt,
Engelund:2012re,
Johansson:2012zv,
Boels:2012ew,
Penante:2014sza,
Brandhuber:2014ica,
Bork:2014eqa,
Wilhelm:2014qua,
Nandan:2014oga,
Loebbert:2015ova,
Bork:2015fla,
Frassek:2015rka,
Boels:2015yna,
Huang:2016bmv} and strong coupling \cite{Maldacena:2010kp,Gao:2013dza} (see \cite{Wilhelm:2016izi} for a review). In particular, already at tree-level, form factors inherited remarkable structures from amplitudes, such as recursion relations, Grassmannian and polytope pictures (see \cite{ArkaniHamed:2012nw} for a review); all these ideas are intimately related to connected prescription for amplitudes. Given the success of twistor-string and scattering-equation-based construction for amplitudes, it is natural to study their applications to form factors especially for ${\cal N}\!=\!4$ SYM, which is the subject of this note.

A $n$-point form factor is given by the overlap of a composite operator with off-shell momentum $q$, and $n$ on-shell states with on-shell momenta $p_1, p_2,\ldots, p_n$:
\begin{align}
{\cal F}_{\cal O} (1,2,\ldots,n) = \int \frac{d^4 x}{ (2\pi)^4}\,e^{-i q\cdot x}\,  \langle 1, \ldots, n | {\cal O}(x) | 0 \rangle = \delta^{(4)} \big(\sum_{i=1}^n p_i - q\big) \langle 1, \ldots, n | {\cal O}(0) | 0  \rangle\,,
\end{align}
where the number of fields in ${\cal O}$ cannot exceed the number of external legs, $n$. In pure Yang-Mills theory, the simplest form factor is that with operator ${\cal O}_{m=2}={\rm Tr} F^2$; it is equivalent to the amplitude with $n$ gluons and a Higgs boson with massive momentum $q$, through the effective coupling $H\,{\rm Tr} F^2$~\cite{Dixon:2004za}. As is well known, in (super-) Yang-Mills theories, it is useful to decompose the operator to chiral and anti-chiral parts, and study the corresponding form factors separately. For example, here we have $F^{\mu \nu}=F_-^{\mu \nu} + F_+^{\mu \nu}$ which are also known as self-dual and anti-self-dual parts, 
and we refer to their form factors as ${\cal F}_{F_-^2,\, n} + {\cal F}_{F_+^2,\, n}$.

The key observation of~\cite{He:2016dol} is to represent the operator effectively as two auxiliary, on-shell legs: the off-shell momentum is $-q=\lambda_x \tilde\lambda_x+\lambda_y \tilde\lambda_y$, and assign two additional punctures $\sigma^{\alpha}_x$ and $\sigma^{\alpha}_y$ for them. It is crucial that the $x,y$ appear together with the $+$ (or $-$ resp.) set in the scattering equations, for the case of ${\cal F}_{F_{\pm}^2, n}$. For general chiral form factors, the prescription given in~\cite{He:2016dol} is to use the following {\it chiral off-shell scattering equations}:
\begin{align}\label{eqFF}
\begin{aligned}
  \tilde\lambda_x - \sum_{i\in +} {\tilde\lambda_i \over (x\,i)} \,=\,0\,,\quad \tilde\lambda_y - \sum_{i\in +} {\tilde\lambda_i \over (y\,i)} \,=\,0\,,\quad \quad
  \tilde\lambda_I - \sum_{i\in +} {\tilde\lambda_i  \over (I\,i)}\,=\,0\,,~~~ I\in -\,;
  \\
  \lambda_i -\sum_{I\in -} {\lambda_I \over (i\,I)}
  - {\lambda_x \over (i\,x)} - {\lambda_y \over (i\,y)}  \,=\,0\,,~~~ i\in +\,,
\end{aligned}
\end{align}
and the anti-chiral case is obtained by parity. Based on \eqref{eqFF}, \eqref{YM} has been generalized to a connected formula for ${\cal F}_{F^2}$~\cite{He:2016dol}, and a particularly nice form follows from fixing the four variables $\sigma^\alpha_x$, $\sigma^\alpha_y$ and pull out their four delta functions:
\begin{align}\label{F2FF}
%
{\cal F}_{F_-^2, n} &= (q^2)^2\,\delta^4(P)
\int {\prod_{a=1}^n d^2\sigma_a \over (12)(23)\cdots(n1)}
\prod_{I\in -} \delta^2\!\left(\tilde\lambda_I-\sum_{i\in +} \frac{\tilde\lambda_i}{(I\,i)}\right)
\prod_{i \in +} \delta^2\!\left(\lambda_i-\sum_{I \in -,x,y} \frac{\lambda_I}{(i\,I)}\right),
\end{align}
where $q^2=\braket{x\,y}[x\,y]$ and $\delta^4(P):=\delta^4\big(\sum_a \lambda_a\tilde\lambda_a + \lambda_x\tilde\lambda_x + \lambda_y\tilde\lambda_y \big)$.\footnote{In this note, the spinor and anti-spinor products are defined as $\braket{i\,j}=\lambda_{i,\alpha}\lambda_j^{\alpha}$,~$[i\,j]
=\tilde\lambda_{i,\dot\alpha}\tilde\lambda_j^{\dot\alpha}$.}
Remarkably the result only depends on $q$ but not on individual momenta of $x,y$. Moreover, the punctures $x,y$ do not enter the Parke-Taylor factor since they do not carry color or little group weight.

In the following we consider extensions of \eqref{F2FF} to more general form factors in ${\cal N}\!=\!4$ SYM.
By supersymmetrizing ${\rm Tr} F_-^2$ one obtains chiral part of the stress-tensor multiplet operator ${\cal T}_2$. It is well known how to supersymmetrize the Yang-Mills formula to ${\cal N}\!=\!4$ SYM: one defines the superamplitude in chiral superspace $(\lambda, \tilde\lambda, \eta)$, with the supercharge $Q^{\alpha,A}=\sum_{a=1}^n \lambda^{\alpha}_a \eta^A_a$; then by inserting fermionic delta functions for $\eta$'s, which are supersymmetric analog of the first half of \eqref{eq} (those for $\tilde\lambda$'s), \eqref{YM} is extended to that for SYM. In a completely parallel way, we will see how to write down the formula for super form factor for ${\cal T}_2$ based on supersymmetrizing \eqref{eqFF}. Note that super amplitudes or super form factors are permutation invariant (independent of the choice of sets $+$ and $-$), which can be made manifest by going to RSV-Witten form.

As a more non-trivial extension, we go beyond bilinear half-BPS operators, and consider bosonic form factor with scalar operators of ${\cal O}_m\equiv\operatorname{Tr}\left[(\phi_{12})^m\right]$. Note that for $m=2$, ${\cal O}_2=\operatorname{Tr}\left[(\phi_{12})^2\right]$ is nothing but the bottom component of ${\cal T}_2$ we studied.  For general $m$, ${\cal O}_m$ is the bottom component of half-BPS operators dual to Kaluza-Klein modes in supergravity~\cite{Penante:2014sza}. We will give a very compact formula for form factors with ${\cal O}_m$ operator which requires external states to be $m$ scalars and $n{-}m$ gluons with arbitrary helicities:
\begin{align}\label{FF-Om}
  {\cal F}_{{\cal O}_m,n} \,\equiv\,
  \delta^4 (P)~\Braket{g(p_1)\cdots\phi_{12}(p_{i_1})\cdots\phi_{12}(p_{i_m})\cdots g(p_n)
  |{\cal O}_m(0)| 0}\,.
\end{align}
A prior it is not obvious what to do for such operators with $m>2$ fields: we could either represent the operator using two on-shell legs as in \eqref{eqFF}, or using $m$ of them. We find very strong evidence for the universality of scattering equations for form factors, \eqref{eqFF}.

\vskip 5mm
{\it Added note}: In the completion of this manuscript, the paper \cite{Brandhuber:2016xue} appeared on the arXiv, which has some overlap with our results for ${\cal T}_2$.

\section{Connected formulas for form factors}
In this section we present the connected formulas for super form factors with the chiral part of the stress-tensor multiplet operator ${\cal T}_2$ in ${\cal N}\!=\!4$ SYM, as well as the bosonic form factor of ${\cal O}_m$ \eqref{FF-Om}, based on the chiral off-shell scattering equations \eqref{eqFF}.

Recall that in on-shell superspace, the super form factor for the operator ${\cal T}_2$ is defined as
\begin{align}\label{super-FF-T2}
  {\mathscr F}_{{\cal T}_2, n} \,:=\, \Braket{\Phi_1\cdots\Phi_n
  |{\cal T}(0)|0},
\end{align}
where external legs $\Phi_i=\Phi_i(p_i,\eta_i)$ are ${\cal N}\!=\!4$ on-shell superfields
\begin{align}\label{N=4-on-shell-superfield}
  \Phi(p,\eta) \,=\, g^+(p) + \eta_A \psi^A(p)
  + {1\over 2!} \eta_{A}\eta_{B}\phi^{AB}(p)
  + {1\over 3!} \eta_{A}\eta_{B}\eta_{C}\, \epsilon^{ABCD} \bar\psi_{D}(p)
  + \eta_{1}\eta_{2}\eta_{3}\eta_{4}\, g^-(p),
\end{align}
where $A=1,2,3,4$ are SU$(4)$ R-symmetry indices and $\phi_{AB}={1\over 2}\epsilon_{ABCD}\phi^{CD}$ are antisymmetric in $A$ and $B$.  Here ${\cal T}_2={\cal T}(x,\theta^+,u)$ is the chiral part of the stress-tensor multiplet operator which has the form in harmonic superspace\footnote{In harmonic superspace:~$\theta^{+a}_\alpha=\theta^A_{\alpha}u^{+a}_A$, $\theta^{-a'}_\alpha=\theta^A_{\alpha}u^{-a'}_A$~with~$a,a'=1,2$, where $(u^{+a}_A, u^{-a'}_A)$ is the normalized harmonic matrix of SU$(4)$.
It is useful to work with harmonic variables for the super form factor of chiral operator. For more details about ${\cal N}\!=\!4$ harmonic superspace, c.f.~\cite{Eden:2011yp, Eden:2011ku}. 

}
\begin{align}\label{T2}
  {\cal T}\big(x,\theta^+,\bar\theta_-=0, u\big) \,=\,& \Tr\left(\phi^{++}\phi^{++}\right)
  + i\,2\sqrt2\,\theta^{+a}_\alpha \Tr\left(\psi^{+\alpha}_a \phi^{++}\right)
  \nonumber
  \\
  &+\theta^{+a}_\alpha\epsilon_{ab}\theta^{+b}_\beta
  \Tr\left(\psi^{+c(\alpha}\psi^{+\beta)}_c - i\sqrt2 F^{\alpha\beta}\phi^{++}\right)
  \nonumber
  \\
  &-\theta^{+a}_\alpha\epsilon^{\alpha\beta}\theta^{+b}_\beta
  \Tr\left(\psi^{+\gamma}_{(a}\psi^{+}_{b)\gamma} - g\sqrt2 \big[ \phi^{+C}_{(a}, \bar\phi_{C\,+b)} \big]\phi^{++}\right)
  \nonumber
  \\
  &-{4\over 3}(\theta^{+})^{3\,a}_{~\alpha}
  \Tr\left(F^\alpha_\beta\psi^{+\beta}_a + ig \big[ \phi^{+B}_{a}, \bar\phi_{BC} \big]\psi^{C\alpha}\right) + {1\over 3}(\theta^{+})^4 {\cal L}(x).
\end{align}
Its $(\theta^+)^0$ component is just the scalar operator, while the $(\theta^+)^4$ component is the chiral form of the ${\cal N}\!=\!4$ on-shell Lagrangian:
\begin{align}\label{}
  {\cal L} \,=\, \Tr\bigg(
  - {1\over 2}F_{\alpha\beta}F^{\alpha\beta}
  + \sqrt2\,g\, \psi^{\alpha A}\big[\phi_{AB}, \psi^B_{\alpha}\big]
  - {1\over 8}g^2 \big[\phi^{AB}, \phi^{CD}\big] \big[\phi_{AB}, \phi_{CD}\big]
  \bigg).
\end{align}
Note that only the chiral half of ${\cal N}\!=\!4$ multiplet, i.e.~$F^{\alpha\beta}$, $\psi^A_{\alpha}$ and $\phi_{AB}$, is needed for ${\cal T}_2$.

\subsection{Connected formula for super form factor}

Although ${\rm Tr}F_-^2$ is contained in ${\cal T}_2$, the natural starting point for supersymmetrization is the form factor with its bottom component, the scalar operator ${\cal O}_2=\operatorname{Tr}\left[(\phi_{12})^2\right]$:
\begin{align}\label{FF-O2}
  {\cal F}_{{\cal O}_2,n} \,\equiv\,
  \Braket{g(p_1)\cdots\phi_{12}(p_i)\cdots\phi_{12}(p_j)\cdots g(p_n)
  |{\cal O}_2(0)| 0}\,.
\end{align}
Once the connected formula for ${\cal F}_{{\cal O}_2}$ is obtained, it is straightforward to obtain the formula for ${\mathscr F}_{{\cal T}_2, n}$ \eqref{super-FF-T2}. The connected formula for \eqref{FF-O2} turns out to be very similar to those for $\Tr F_-^2$ and for tree amplitudes:
\begin{align}\label{CHY-FF-O2}
  {\cal F}^{(k)}_{{\cal O}_2, n}
  \,=\, \delta^4\left(P\right)
  \int d\mu^{(k)}_n\,
  {{\cal I}_2(i,j; x,y) \over (1\,2) \cdots (n\,1)}
\end{align}
where the measure in N${}^{k{-}2}$MHV sector (note $|-|=k$ and $|+|=n-k$) is defined as
\begin{align}\label{msasure-b}
  d\mu^{(k)}_n \,=\, \left(\prod_{i=1}^{n} d^{2}\sigma_i\right)\,
  \prod_{I\in -}^n \delta^{2}
  \left(\tilde\lambda_I - \sum_{i\in +}{\tilde\lambda_i \over (I\,i)}\right)
  \prod_{i\in +} \delta^{2}
  \left(\lambda_i - \sum_{I\in -}{\lambda_I \over (i\,I)}
   - {\lambda_x \over (i\,x)} - {\lambda_y \over (i\,y)}\right),
\end{align}
and it is exactly the same measure as that for the form factor of $\Tr F_-^2$ in \eqref{F2FF}. The only difference is that there is an additional factor that depends on $x,y$ and $i, j$ (positions of external scalars)
\begin{align}\label{I2}
  {\cal I}_2(i,j; x,y) \,=\, -{\braket{xy}^2\,(i\,j)^2
  \over (i\,x)^2(i\,y)^2 (j\,x)^2(j\,y)^2}.
\end{align}
A remarkable property is that the function ${\cal I}_2(i,j; x,y)$ is independent of information of external gluon legs such it has the exact same form for all helicity sectors. 

Now we can generalize our formula \eqref{CHY-FF-O2} along two directions\,--\,one is to supersymmetrize it to for the super form factor \eqref{super-FF-T2}, and the other is to extend it to bosonic form factors \eqref{FF-Om}.

Let us address the super form factor first. In addition to the bosonic part, \eqref{eqFF}, ${\cal N}\!=\!4$ SUSY also requires fermionic delta functions which can be viewed as the super partner of scattering equations.
Our notation is that $\gamma^{\alpha}_{+a}$ stands for the Grassmann variable conjugate to $\theta^{+a}_{\alpha}$ for ${\cal T}_2$ (there is no $\gamma^-$ because we only consider the chiral part of the stress tensor), and we introduce
\begin{align}\label{}
  \eta_{\pm a,i} \,:=\, \bar u^A_{\pm a}\eta_{A,i}
\end{align}
for each on-shell external leg $i$.
Thus the supermomentum $Q^\alpha_A$ can be written as:
\begin{align}\label{}
  Q^{\alpha}_{+a} \,=\, \gamma_{+a}^\alpha - \sum_{i=1}^n \lambda_i^\alpha\eta_{+a,i}\,,
  \qquad
  Q^{\alpha}_{-a} \,=\, \sum_{i=1}^n \lambda_i^\alpha\eta_{-a,i}\,.
\end{align}
Note that we need to split the indices of $\gamma_+$ into two parts by projecting it along two directions,
\begin{align}\label{}
  \eta_{+a,x} \,\equiv\, {\lambda_{y,\alpha}\gamma_{+a}^{\alpha} \over \braket{y\,x}},
  \qquad
  \eta_{+a,y} \,\equiv\, {\lambda_{x,\alpha}\gamma_{+a}^{\alpha} \over \braket{x\,y}}.
\end{align}
The fermonic delta functions for $\eta$'s then take the same form as the delta functions for $\tilde\lambda$ (the first line of \eqref{eqFF}). By combining the bosonic and fermionic part together, we obtain the supersymmetric measure in the $k$-sector of the connected formula:
\begin{align}\label{}
  d\mu^{(k)}_{n,\,{\cal N}=4} \,=\, &\left(\prod_{i=1}^{n} d^{2}\sigma_i\right)\,
  \prod_{I=1}^k \delta^{2}
  \left(\tilde\lambda_I - \sum_{i=k+1}^n {\tilde\lambda_i \over (I\,i)}\right)
  \prod_{i=k+1}^n \delta^{2}
  \left(\lambda_i - \sum_{I=1}^k {\lambda_I \over (i\,I)}
  - {\lambda_x \over (i\,x)} - {\lambda_y \over (i\,y)}\right)
  \nonumber
  \\
  &\times
  \prod_{I=1}^{k}
  \delta^{0|2}\left(\eta_{+,I} - \sum_{i=k+1}^n {\eta_{+,i} \over (I\,i)} \right)
  \delta^{0|2}\left(\eta_{-,I} - \sum_{i=k+1}^n {\eta_{-,i} \over (I\,i)} \right)
  \delta^{0|2}\left({\lambda_{y,\alpha}\gamma_{+}^{\alpha} \over \braket{y\,x}}
  - \sum_{i=k+1}^n {\eta_{+,i} \over (x\,i)} \right)
  \nonumber
  \\
  &\times
  \delta^{0|2}\left({\lambda_{x,\alpha}\gamma_{+}^{\alpha} \over \braket{x\,y}}
  - \sum_{i=k+1}^n {\eta_{+,i} \over (y\,i)} \right)
  \delta^{0|2}\left( \sum_{i=k+1}^n {\eta_{-,i} \over (x\,i)} \right)
  \delta^{0|2}\left( \sum_{i=k+1}^n {\eta_{-,i} \over (y\,i)} \right)\,,
\end{align}
where we take  $-=\{1,\ldots,k\}$ and $+=\{k+1,\ldots,n\}$ to make it completely explicit.

To write down the complete formula, let us take a closer look at formula \eqref{CHY-FF-O2} again. An interesting observation is that the function ${\cal I}_2(i,j; x,y)$ \eqref{I2} can be understood as coming from the $\eta$-projection when we evaluate the fermonic delta functions of the supersymmetric formula. To conclude, all we need for the super form factor \eqref{super-FF-T2} is simply the Parke-Taylor factor!
\begin{align}\label{CHY-FF-T2}
  {\mathscr F}^{(k)}_{{\cal T}_2, n}
  \,=\, \delta^4\left(P\right)
  \int d\mu^{(k)}_{n,\,{\cal N}=4}\,
  {\braket{x\,y}^4 \over (1\,2) \cdots (n\,1)}.
\end{align}
In \eqref{CHY-FF-T2}, one has fixed the four variables $\sigma^\alpha_x$, $\sigma^\alpha_y$ and pull out their four delta functions as $\delta^4(P)$ when removing $\operatorname{GL}(2,{\mathbb C})$ redundancy. As we have mentioned, \eqref{CHY-FF-T2} is very similar to that for superamplitude in ${\cal N}\!=\!4$ SYM\,--\,the only modification is to include two on-shell legs $x,y$ in a supersymmetric way, to represent the operator ${\cal T}_2$.

As discussed in~\cite{He:2016vfi}, it is straightforward to translate the formula to a manifestly permutation-invariant form, similar to the RSV-Witten formula for SYM amplitudes~\cite{Witten:2003nn, Roiban:2004yf}. By applying the same procedure to \eqref{CHY-FF-T2}, we obtain an equivalent formula for the super form factor with ${\cal T}_2$
\begin{align}\label{RSV-FF-T2}
  {\mathscr F}^{(k)}_{{\cal T}_2, n}
  \,=\, \int d\widetilde{\mu}^{(k)}_{n,\,{\cal N}=4}\,
  {\braket{x\,y}^2 \over (1\,2) \cdots (n\,1)}
\end{align}
with the supersymmetric measure similar to the RSV-Witten form
\begin{align}\label{}
  d\widetilde{\mu}^{(k)}_{n,\, {\cal N}=4}
  \,:=\, {d^{2n+4}\sigma \over \operatorname{vol} {\rm GL}(2,{\mathbb C})}
  & \prod_{m=0}^{k+1} \delta^{2|2}
  \left( \big( t_x\sigma_x^m\tilde\lambda_x
  +  t_y \sigma_y^m\tilde\lambda_y|0\big)
  +\sum_{i=1}^n  t_i \sigma_i^m \big(\tilde\lambda_i|\eta_{-,i}\big) \right)
  \nonumber\\
  \times&
  \prod_{m=0}^{k+1} \delta^{2|2}
  \left( t_x \sigma_x^m\big(\tilde\lambda_x|\eta_{+,x}\big)
  + t_y \sigma_y^m \big(\tilde\lambda_y|\eta_{+,y}\big)
  +\sum_{i=1}^n  t_i \sigma_i^m (\tilde\lambda_i|\eta_{+,i}) \right)
  \nonumber\\
  \times&
  \int d^{2k+4}\rho \prod_{i=1,\ldots,n,x,y}
  \delta^{2} \left(  t_i\sum_{m=0}^{k+1} \rho_m \sigma_i^m
  - \lambda_i \right),
\end{align}
where $(a\,b):=(\sigma_a - \sigma_b)/(t_at_b)$ and $d^{2n+4}\sigma := \prod_{i=1}^{n} d^2\sigma_i\,d^2\sigma_x d^2\sigma_y$ with $d^2\sigma_a = d\sigma_a dt_a/t_a^3$.

\subsection{Connected formula for bosonic form factors}

Now we consider the bonsonic form factor with the operator ${\cal O}_m$ \eqref{FF-Om}. The key observation is that no matter how many fields in the operator, we only need two additional legs/punctures $x,y$, and $d\mu_n^{(k)}$ (including delta functions imposing \eqref{eqFF}) is exactly the same as in the $m=2$ case.

For ${\cal O}_m$, all we need is simple modification of Parke-Taylor factor which takes into account the positions of the $m$ on-shell scalars. This supports our general conjecture that all the information of the operator and external states is encoded in the integrand of connected formula. In other words, what we find is that similar to $m=2$ case, now we have a function ${\cal I}_{m}$ that incorporates the information of ${\cal O}_m$ and $m$ on-shell scalars. Our proposal is the following formula
\begin{align}\label{CHY-FF-Om}
  {\cal F}^{(k)}_{{\cal O}_m, n}
  \,=\, \delta^4\left(P\right)
  \int d\mu^{(k)}_{n}\,
  {{\cal I}_m(i_1,\ldots,i_m; x,y) \over (1\,2) \cdots (n\,1)}
\end{align}
with the measure $d\mu^{(k)}_n$ is exactly same with the one for ${\cal O}_2$ \eqref{msasure-b}, and
\begin{align}\label{}
  {\cal I}_m(i_1,\ldots,i_m; x,y) \,=\, \braket{x\,y}^m\,{(i_1\,i_2)\cdots(i_m\,i_1)
  \over \prod_{\alpha=1}^m (i_\alpha\,x)^2(i_\alpha\, y)^2},
\end{align}
which goes back to \eqref{I2} when $m=2$.
Starting from \eqref{CHY-FF-Om}, it is also possible to construct the connected formula for the supersymmetric operator of form $\displaystyle
  {\cal T}_m\big(x,\theta^+\big) \,:=\, \Tr\left[\big(W^{++}(x,\theta^+)\big)^m\right]
$
which is a generalization of ${\cal T}_2$, because ${\cal O}_m$ is the bottom component of ${\cal T}_m$. The validity for both formulas \eqref{CHY-FF-T2} and \eqref{CHY-FF-Om} may be established by careful study of their behavior under factorization and soft limits, which are almost identical to the amplitude case~\cite{He:2016dol}. We leave the proof for future works, and content ourselves here with examples of explicit calculations.

\subsection{Examples}

As consistency checks for both formulas \eqref{CHY-FF-T2} and \eqref{CHY-FF-Om}, we evaluate them and compare with known results for some form factors. These include MHV and maximally-non-MHV super form factors for all multiplicities, as well as some more complicated component form factors.

\subsubsection{MHV}
First we consider the MHV sector which corresponds to $k=0$.
The MHV scattering equations have only a unique solution after fixing $\operatorname{GL}(2)$-redundancy \cite{Cachazo:2013iaa,Dolan:2014ega, Weinzierl:2014ava}.
Here it is convenient to fix the locations for two ``additional'' punctures $\sigma_{x}$, $\sigma_{y}$ such that $(x\,y)=1$.\footnote{To be precise, for example, this can be realized via fixing $\sigma_x^{\alpha}=(1,0)$,~$\sigma_y^{\alpha}=(0,1)$.}
Thus the unique solutions for the MHV equations is given by:
\begin{align}\label{MHV-sol}
  (i\,x) \,=\, {\braket{x\,y}\over\braket{i\,y}}, \quad
  (i\,y) \,=\, {\braket{x\,y}\over\braket{x\,i}}, \quad
  (i\,j) \,=\, {\braket{x\,y}^3\braket{i\,j} \over
  \braket{x\,i}\braket{y\,i} \braket{x\,j}\braket{y\,j}}.
\end{align}
Here $(i\,j)$ is nothing but the reciprocal of link variable,~$c_{i,j}:=1/(i\,j)$.

Note that the integral measure can be written nicely in terms of the $2n$ 2-brackets $(i x)$, $(i y)$
\begin{align}\label{measure-jac}
  \prod_{i=1}^n {d^2\sigma_i} \,=\, \prod_{i=1}^n d(ix)\,d(iy),
\end{align}
and all the delta functions can be rewritten in terms of them,
\begin{align}\label{delta-jac}
  \delta^{2} \left(\lambda_i - {\lambda_x \over (i\,x)} - {\lambda_y \over (i\,y)} \right)
  \,&=\, {\braket{xy}^3 \over \braket{ix}^2\braket{iy}^2}\,
  \delta\!\left( (ix) - {\braket{i\,y}\over \braket{x\,y}}\right)\,
  \delta\!\left( (iy) - {\braket{x\,i}\over \braket{x\,y}}\right).
\end{align}

Plugging the solution \eqref{MHV-sol} and the Jacobian for measure and delta-functions, \eqref{measure-jac}, \eqref{delta-jac}, into \eqref{CHY-FF-Om} and \eqref{CHY-FF-T2} respectively, we obtain immediately the correct result for the bosonic form factor
\begin{align}\label{MHV-FF-O2}
  {\cal F}_{{\cal O}_m, n}^\text{MHV}
  \,=\, \delta^4(P){\braket{i_1\,i_2}\braket{i_2\,i_3}\cdots\braket{i_m\,i_1}
  \over \braket{1\,2}\braket{2\,3}\cdots\braket{n1}},
\end{align}
and similarly for the super form factor
\begin{align}\label{MHV-FF-T2}
  {\mathscr F}^\text{MHV}_{{\cal T}_2, n}
  \,=\,
  {\displaystyle\delta^4(P)\,
  \delta^4\!\left(\gamma_{+} \,-\, \sum_{i=1}^n\lambda_{i}\eta_{+,i}\right)
  \delta^4\!\left(\sum_{i=1}^n\lambda_{i}\eta_{-,i}\right)
  \over \braket{1\,2}\braket{2\,3}\cdots\braket{n1}},
\end{align}
where in the fermonic delta functions we have pulled out two overall factors of $1/\langle x\,y \rangle^2$:
\begin{align}\label{}
&&\delta^2\left({\lambda_{y,\alpha}\gamma_{+}^{\alpha} \over \braket{y\,x}}
  - \sum_{i=1}^n {\eta_{+,i} \over (x\,i)} \right)
  \delta^2\left({\lambda_{x,\alpha}\gamma_{+}^{\alpha} \over \braket{x\,y}}
  - \sum_{i=1}^n {\eta_{+,i} \over (y\,i)} \right)
  \,=\, {1\over \braket{x\,y}^2} \delta^4\left(\gamma_{+} \,-\, \sum_{i=1}^n\lambda_{i}\eta_{+,i}\right),\\\nonumber
&&\delta^2\left( \sum_{i=1}^n {\eta_{-,i} \over (x\,i)} \right)
  \delta^2\left( \sum_{i=1}^n {\eta_{-,i} \over (y\,i)} \right)
  \,=\, {1\over \braket{x\,y}^2} \delta^4\left(\sum_{i=1}^n\lambda_{i}\eta_{-,i}\right).
\end{align}

\subsubsection{Maximally non-MHV}
Similarly, it is also easy to get the $\overline{\text{MHV}}$ (usually called maximally non-MHV) form factor with ${\cal T}_2$ by simply flipping the helicities of spinors for the MHV formula, i.e.,
\begin{align}\label{super-CHY-FF-T2}
  {\mathscr F}^{\overline{\rm MHV}}_{{\cal T}_2, n}
  \,=\,
  \int d\mu_n\,{\braket{x\,y}^2[x\,y]^2 \over (1\,2) \cdots (n\,1)}
\end{align}
with
\begin{align}\label{}
  d\mu_n = \left(\prod_{i=1}^{n} d^{2}\sigma_i\right)
  &\prod_{I=1}^n \delta^{2}
  \left(\tilde\lambda_I - {\tilde\lambda_x \over (I\,x)}
  - {\tilde\lambda_y \over (I\,y)}\right)
  \prod_{I=1}^{n}
  \delta^{0|2}\left(\eta_{+,I} - {\eta_{+,x} \over (I\,x)} - {\eta_{+,y} \over (I\,y)} \right)
  \delta^{0|2}\big(\eta_{-,I} \big).
\end{align}
The solution for this case is nothing but solution \eqref{MHV-sol} by replacing $\braket{~~}\to [~~]$.
Imposing the unique solution and the Jacobian for the bosonic delta-functions, one can immediately obtain
\begin{align}\label{anti-MHV-FF-T2}
  {\mathscr F}^{\overline{\rm MHV}}_{{\cal T}_2, n}
  \,=\, \delta^4(P) {q^4 \over [1\,2][2\,3] \cdots [n\,1]}
  \prod_{I=1}^{n}
  \delta^{0|2}\left(\eta_{+,I} - {[I\,y] \over [x\,y]}\eta_{+,x}
  - {[x\,I] \over [x\,y]} \eta_{+,y} \right)
  \delta^{0|2}\big(\eta_{-,I} \big).
\end{align}
Its $(\eta_x)^0(\eta_y)^0$ component is the famous Sudakov form factor.

\subsubsection{NMHV and NNMHV form factors}

We have evaluated our formulas for more involved examples where one needs to sum over multiple solutions to scattering equations. For the form factor of ${\cal O}_2$, we have evaluated it numerically for four-point NMHV, five-point NMHV and NNMHV, as well as six-point NMHV and NNMHV cases. This is done by first solving \eqref{eqFF} and then summing over the resulting 4, 11, 11, 26 and 66 solutions respectively. The result can also be computed from MHV rules or recursion relations as in~\cite{Brandhuber:2011tv}, and in all these cases we find perfect agreement. These checks provide very strong evidence that our formulas \eqref{CHY-FF-O2}  and \eqref{CHY-FF-T2} arecorrect.

In addition, we have evaluated the following two NMHV form factors analytically, and verified that they give correct results~\cite{Brandhuber:2011tv,Penante:2014sza}:
\begin{align}
  \label{NMHV-4pt-O2}
  {F}_{{\cal O}_2}
  \big(\phi_{12}, \phi_{12}, g^-, g^+; q\big)
  \,&=\,
  {1 \over \braket{41}[23]s_{34}}
  \left( {\braket{14}\braket{23}[24]^2 \over s_{234}}
  + {[14][23]\braket{13}^2 \over s_{134}} - \braket{13}[24] \right),
  \\
  \label{NMHV-4pt-O3}
  {F}_{{\cal O}_3}
  \big(\phi_{12}, \phi_{12}, \phi_{12}, g^-; q\big)
  \,&=\, {[31] \over [34][41]}.
\end{align}
Here we need the six-point NMHV scattering equations:
\begin{align}\label{6pt-SE}
  \tilde\lambda_I - \sum_{i} {\tilde\lambda_i \over (I\,i)} = 0,\quad
  \tilde\lambda_x - \sum_{i} {\tilde\lambda_i \over (x\,i)} = 0,\quad
  \tilde\lambda_y - \sum_{i} {\tilde\lambda_i \over (y\,i)} = 0;\quad
  \lambda_i - \sum_{J=I,x,y}{\lambda_J \over (i\,J)} = 0,
\end{align}
where $I=3$, $i=1,2,4$ for the first form factor, \eqref{NMHV-4pt-O2}, and $I=4$, $i=1,2,3$ for the second, \eqref{NMHV-4pt-O3}.
In this special case, one can evaluate the formula \eqref{CHY-FF-Om} by following the same procedure of~\cite{ArkaniHamed:2009si, ArkaniHamed:2009dn}. One first introduces 9 link variables $c_{I,i}:=1/(I\,i)$ and treat \eqref{6pt-SE} as 8 linear equations for these variables, subject to an additional ``Veronese constraint". The formula \eqref{CHY-FF-Om} thus turns to a one-dimensional contour integral, and by residue theorem, it become sum over residues which coincide with \eqref{NMHV-4pt-O2} and \eqref{NMHV-4pt-O3} respectively. This serves as not only further consistency checks, but also examples for evaluating our formulas using link variables in general.

\section{Discussions and outlook}

In this note we have initiated the investigation on connected prescription for tree-level form factors in ${\cal N}=4$ SYM, which can be viewed as natural generalizations of \eqref{F2FF} proposed in \cite{He:2016dol}. An example we considered is the chiral part of stress-tensor multiplet operator ${\cal T}_2$, which has ${\cal O}_2=\operatorname{Tr}\left[(\phi_{12})^2\right]$ as its bottom component. The formula, \eqref{CHY-FF-T2}, is the supersymmetrization of \eqref{F2FF}, which one can also naturally written in the RSV-Witten form. More interestingly, we studied generalizations to half-BPS operator ${\cal O}_m$ for arbitrary $m$ which are generally no longer bilinear, and obtained again very compact formula \eqref{CHY-FF-Om}. Our result strongly support the universality of the scattering equations \eqref{eqFF} for chiral form factors.

In~\cite{ArkaniHamed:2009dg}, an interesting ``duality" has been proposed, which relates connected formula and $G(k,n)$ Grassmannian formula for amplitudes in ${\cal N}\!=\!4$ SYM. By repeated use of the Global residue theorem, one can rewrite the connected formula as particular sums of residues of the Grassmannian contour integral, which give tree amplitudes in the so-called link representation. The same conclusion holds for our connected formula for form factor, as we have seen in the two NMHV examples above. In general, the formula must admit link representation which are given by particular contours for the integrals over $G(k+2, n+2)$ (or $G(k,n+2)$ for anti-chiral part), where $x,y$ do not enter the cyclic measure. This agrees with the Grassmannian formula for ${\cal F}_{{\cal T}_2}$ in~\cite{Frassek:2015rka}, and we expect a careful investigation can pick out the precise contour that gives the tree form factor. Generic residues of the Grassmannian contour integral correspond to on-shell diagrams, which have manifest symmetry properties and significance for loops. It would be very tempting to see if the connected formula shed new lights on these ideas for form factors.

We expect that our results can be generalized to form factors for general operators in ${\cal N}\!=\!4$ SYM at tree level. Very recently, a closed formula for all MHV form factors with most general operators was given in~\cite{Koster:2016loo}, and it is likely that with connected prescription it can be generalized to any N${}^k$MHV sector. It would also be very interesting to compare with these recent works on SYM form factors using twistor space method~\cite{Koster:2016ebi, Chicherin:2016soh}, which can help us understand better off-shell quantities in twistor space. It is plausible that the idea of connected prescriptions and twistor strings can be applied to correlation functions in SYM (see ~\cite{Chicherin:2014uca}). Last but not least, most results have been restricted to four dimensions, but scattering-equation-based construction can be useful for studying form factors in general dimensions as well. Given the success of scattering equations at loop level, a natural next step would be applying them to loop-level form factors.

\section*{Acknowledgments}
The authors thank Reza Doobary, Gang Yang and Yong Zhang for discussions.
ZL~is supported by the ``Fonds Sp\'{e}cial de Recherche'' (FSR) of the UCLouvain.


\begin{thebibliography}{100}
\itemsep=-0.55mm
\small

\bibitem{Witten:2003nn}
E.~Witten,
{\it Perturbative gauge theory as a string theory in twistor space},
{Commun.\ Math.\ Phys.\  {\bf 252} (2004) 189}
[\href{http://arxiv.org/abs/hep-th/0312171}
{\tt hep-th/0312171}]

\bibitem{Roiban:2004yf}
R.~Roiban, M.~Spradlin and A.~Volovich,
{\it On the tree level S-matrix of Yang-Mills theory},
{Phys.\ Rev.\ D {\bf 70} (2004) 026009}
[\href{http://arxiv.org/abs/hep-th/0403190}
{\tt hep-th/0403190}]

\bibitem{Cachazo:2013gna}
F.~Cachazo, S.~He and E.~Y.~Yuan,
{\it Scattering equations and Kawai-Lewellen-Tye orthogonality},
{Phys.\ Rev.\ D {\bf 90} (2014) 065001}
[\href{http://arxiv.org/abs/1306.6575}
{\tt arXiv:1306.6575}]

\bibitem{Cachazo:2013hca}
F.~Cachazo, S.~He and E.~Y.~Yuan,
{\it Scattering of Massless Particles in Arbitrary Dimensions},
{Phys.\ Rev.\ Lett.\  {\bf 113} (2014) 171601}
[\href{http://arxiv.org/abs/1307.2199}
{\tt arXiv:1307.2199}]

\bibitem{Cachazo:2013iea}
F.~Cachazo, S.~He and E.~Y.~Yuan,
{\it Scattering of Massless Particles: Scalars, Gluons and Gravitons},
{JHEP {\bf 1407} (2014) 033}
[\href{http://arxiv.org/abs/1309.0885}
{\tt arXiv:1309.0885}]

\bibitem{Cachazo:2014nsa}
F.~Cachazo, S.~He and E.~Y.~Yuan,
{\it Einstein-Yang-Mills Scattering Amplitudes From Scattering Equations},
{JHEP {\bf 1501} (2015) 121}
[\href{http://arxiv.org/abs/1409.8256}
{\tt arXiv:1409.8256}]

\bibitem{Cachazo:2014xea}
F.~Cachazo, S.~He and E.~Y.~Yuan,
{\it Scattering Equations and Matrices:~From Einstein To Yang-Mills, DBI and NLSM},
{JHEP {\bf 1507} (2015) 149}
[\href{http://arxiv.org/abs/1412.3479}
{\tt arXiv:1412.3479}]

\bibitem{Geyer:2016wjx}
Y.~Geyer, L.~Mason, R.~Monteiro and P.~Tourkine,
{\it Two-Loop Scattering Amplitudes from the Riemann Sphere},
\href{http://arxiv.org/abs/1607.08887}
{\tt arXiv:1607.08887}

\bibitem{Geyer:2015bja}
Y.~Geyer, L.~Mason, R.~Monteiro and P.~Tourkine,
{\it Loop Integrands for Scattering Amplitudes from the Riemann Sphere},
{Phys.\ Rev.\ Lett.\  {\bf 115} (2015) 121603}
[\href{http://arxiv.org/abs/1507.00321}
{\tt arXiv:1507.00321}]

\bibitem{Geyer:2015jch}
Y.~Geyer, L.~Mason, R.~Monteiro and P.~Tourkine,
{\it One-loop amplitudes on the Riemann sphere},
{JHEP {\bf 1603} (2016) 114}
[\href{http://arxiv.org/abs/1511.06315}
{\tt arXiv:1511.06315}]

\bibitem{Cachazo:2015aol}
F.~Cachazo, S.~He and E.~Y.~Yuan,
{\it One-Loop Corrections from Higher Dimensional Tree Amplitudes},
{JHEP {\bf 1608} (2016) 008}
[\href{http://arxiv.org/abs/1512.05001}
{\tt arXiv:1512.05001}]

\bibitem{Mason:2013sva}
L.~Mason and D.~Skinner,
{\it Ambitwistor strings and the scattering equations},
{JHEP {\bf 1407} (2014) 048}
[\href{http://arxiv.org/abs/1311.2564}
{\tt arXiv:1311.2564}]

\bibitem{Cachazo:2012da}
F.~Cachazo and Y.~Geyer,
{\it A `Twistor String' Inspired Formula For Tree-Level Scattering Amplitudes in N=8 SUGRA},
\href{http://arxiv.org/abs/1206.6511}
{\tt arXiv:1206.6511}

\bibitem{Cachazo:2012kg}
F.~Cachazo and D.~Skinner,
{\it Gravity from Rational Curves in Twistor Space},
{Phys.\ Rev.\ Lett.\  {\bf 110} (2013) 161301}
[\href{http://arxiv.org/abs/1207.0741}
{\tt arXiv:1207.0741}]

\bibitem{Cachazo:2012pz}
F.~Cachazo, L.~Mason and D.~Skinner,
{\it Gravity in Twistor Space and its Grassmannian Formulation},
{SIGMA {\bf 10} (2014) 051}
[\href{http://arxiv.org/abs/1207.4712}
{\tt arXiv:1207.4712}]

\bibitem{Skinner:2013xp}
D.~Skinner,
{\it Twistor Strings for N=8 Supergravity},
\href{http://arxiv.org/abs/1301.0868}
{\tt arXiv:1301.0868}

\bibitem{Cachazo:2013iaa}
F.~Cachazo, S.~He and E.~Y.~Yuan,
{\it Scattering in Three Dimensions from Rational Maps},
{JHEP {\bf 1310} (2013) 141}
[\href{http://arxiv.org/abs/1306.2962}
{\tt arXiv:1306.2962}]

\bibitem{Geyer:2014fka}
Y.~Geyer, A.~E.~Lipstein and L.~J.~Mason,
{\it Ambitwistor Strings in Four Dimensions},
{Phys.\ Rev.\ Lett.\  {\bf 113} (2014) 081602}
[\href{http://arxiv.org/abs/1404.6219}
{\tt arXiv:1404.6219}]

\bibitem{Adamo:2015gia}
T.~Adamo, E.~Casali, K.~A.~Roehrig and D.~Skinner,
{\it On tree amplitudes of supersymmetric Einstein-Yang-Mills theory},
{JHEP {\bf 1512} (2015) 177}
[\href{http://arxiv.org/abs/1507.02207}
{\tt arXiv:1507.02207}]

\bibitem{Cachazo:2016sdc}
F.~Cachazo and G.~Zhang,
{\it Minimal Basis in Four Dimensions and Scalar Blocks},
\href{http://arxiv.org/abs/1601.06305}
{\tt arXiv:1601.06305}

\bibitem{He:2016vfi}
S.~He, Z.~Liu and J.-B.~Wu,
{\it Scattering Equations, Twistor-string Formulas and Double-soft Limits in Four Dimensions},
{JHEP {\bf 1607} (2016) 060}
[\href{http://arxiv.org/abs/1604.02834}
{\tt arXiv:1604.02834}]

\bibitem{Cachazo:2016njl}
F.~Cachazo, P.~Cha and S.~Mizera,
{\it Extensions of Theories from Soft Limits},
{JHEP {\bf 1606} (2016) 170}
[\href{http://arxiv.org/abs/1604.03893}
{\tt arXiv:1604.03893}]

\bibitem{He:2016dol}
S.~He and Y.~Zhang,
{\it Connected formulas for amplitudes in standard model},
\href{http://arxiv.org/abs/1607.02843}
{\tt arXiv:1607.02843}
%

\bibitem{Brandhuber:2010ad}
A.~Brandhuber, B.~Spence, G.~Travaglini and G.~Yang,
{\it Form Factors in ${\cal N}\!=\!4$ Super Yang-Mills and Periodic Wilson Loops},
{JHEP {\bf 1101} (2011) 134}
[\href{http://arxiv.org/abs/1011.1899}
{\tt arXiv:1011.1899}]

\bibitem{Bork:2010wf}
L.~V.~Bork, D.~I.~Kazakov and G.~S.~Vartanov,
{\it On form factors in ${\cal N}\!=\!4$ SYM},
{JHEP {\bf 1102} (2011) 063}
[\href{http://arxiv.org/abs/1011.2440}
{\tt arXiv:1011.2440}]

\bibitem{Brandhuber:2011tv}
A.~Brandhuber, O.~Gurdogan, R.~Mooney, G.~Travaglini and G.~Yang,
{\it Harmony of Super Form Factors},
{JHEP {\bf 1110} (2011) 046}
[\href{http://arxiv.org/abs/1107.5067}
{\tt arXiv:1107.5067}]

\bibitem{Penante:2014sza}
B.~Penante, B.~Spence, G.~Travaglini and C.~Wen,
{\it On super form factors of half-BPS operators in ${\cal N}\!=\!4$ super Yang-Mills},
{JHEP {\bf 1404} (2014) 083}
[\href{http://arxiv.org/abs/1402.1300}
{\tt arXiv:1402.1300}]

\bibitem{Bork:2011cj}
L.~V.~Bork, D.~I.~Kazakov and G.~S.~Vartanov,
{\it On MHV Form Factors in Superspace for $\mathcal{N}\!=\!4$ SYM Theory},
{JHEP {\bf 1110} (2011) 133}
[\href{http://arxiv.org/abs/1107.5551}
{\tt arXiv:1107.5551}]

\bibitem{Henn:2011by}
J.~M.~Henn, S.~Moch and S.~G.~Naculich,
{\it Form factors and scattering amplitudes in ${\cal N}\!=\!4$ SYM in dimensional and massive regularizations},
{JHEP {\bf 1112} (2011) 024}
[\href{http://arxiv.org/abs/1109.5057}
{\tt arXiv:1109.5057}]

\bibitem{Gehrmann:2011xn}
T.~Gehrmann, J.~M.~Henn and T.~Huber,
{\it The three-loop form factor in ${\cal N}\!=\!4$ super Yang-Mills},
{JHEP {\bf 1203} (2012) 101}
[\href{http://arxiv.org/abs/1112.4524}
{\tt arXiv:1112.4524}]

\bibitem{Brandhuber:2012vm}
A.~Brandhuber, G.~Travaglini and G.~Yang,
{\it Analytic two-loop form factors in ${\cal N}\!=\!4$ SYM},
{JHEP {\bf 1205} (2012) 082}
[\href{http://arxiv.org/abs/1201.4170}
{\tt arXiv:1201.4170}]

\bibitem{Bork:2012tt}
L.~V.~Bork,
{\it On NMHV form factors in ${\cal N}\!=\!4$ SYM theory from generalized unitarity},
{JHEP {\bf 1301} (2013) 049}
[\href{http://arxiv.org/abs/1203.2596}
{\tt arXiv:1203.2596}]

\bibitem{Engelund:2012re}
O.~T.~Engelund and R.~Roiban,
{\it Correlation functions of local composite operators from generalized unitarity},
{JHEP {\bf 1303} (2013) 172}
[\href{http://arxiv.org/abs/1209.0227}
{\tt arXiv:1209.0227}]

\bibitem{Johansson:2012zv}
H.~Johansson, D.~A.~Kosower and K.~J.~Larsen,
{\it Two-Loop Maximal Unitarity with External Masses},
{Phys.\ Rev.\ D {\bf 87} (2013) 025030}
[\href{http://arxiv.org/abs/1208.1754}
{\tt arXiv:1208.1754}]

\bibitem{Boels:2012ew}
R.~H.~Boels, B.~A.~Kniehl, O.~V.~Tarasov and G.~Yang,
{\it Color-kinematic Duality for Form Factors},
{JHEP {\bf 1302} (2013) 063}
[\href{http://arxiv.org/abs/1211.7028}
{\tt arXiv:1211.7028}]

\bibitem{Brandhuber:2014ica}
A.~Brandhuber, B.~Penante, G.~Travaglini and C.~Wen,
{\it The last of the simple remainders},
{JHEP {\bf 1408} (2014) 100}
[\href{http://arxiv.org/abs/1406.1443}
{\tt arXiv:1406.1443}]

\bibitem{Bork:2014eqa}
L.~V.~Bork,
{\it On form factors in ${\cal N}\!=\!4$ SYM theory and polytopes},
{JHEP {\bf 1412} (2014) 111}
[\href{http://arxiv.org/abs/1407.5568}
{\tt arXiv:1407.5568}]

\bibitem{Wilhelm:2014qua}
M.~Wilhelm,
{\it Amplitudes, Form Factors and the Dilatation Operator in $\mathcal{N}\!=\!4$ SYM Theory},
{JHEP {\bf 1502} (2015) 149}
[\href{http://arxiv.org/abs/1410.6309}
{\tt arXiv:1410.6309}]

\bibitem{Nandan:2014oga}
D.~Nandan, C.~Sieg, M.~Wilhelm and G.~Yang,
{\it Cutting through form factors and cross sections of non-protected operators in $\mathcal{N}\!=\!4 $ SYM},
{JHEP {\bf 1506} (2015) 156}
[\href{http://arxiv.org/abs/1410.8485}
{\tt arXiv:1410.8485}]

\bibitem{Loebbert:2015ova}
F.~Loebbert, D.~Nandan, C.~Sieg, M.~Wilhelm and G.~Yang,
{\it On-Shell Methods for the Two-Loop Dilatation Operator and Finite Remainders},
{JHEP {\bf 1510} (2015) 012}
[\href{http://arxiv.org/abs/1504.06323}
{\tt arXiv:1504.06323}]


\bibitem{Frassek:2015rka}
R.~Frassek, D.~Meidinger, D.~Nandan and M.~Wilhelm,
{\it On-shell diagrams, Gra\ss{}mannians and integrability for form factors},
{JHEP {\bf 1601} (2016) 182}
[\href{http://arxiv.org/abs/1506.08192}
{\tt arXiv:1506.08192}]

\bibitem{Bork:2015fla}
L.~V.~Bork and A.~I.~Onishchenko,
{\it On soft theorems and form factors in $\mathcal{N}\!=\!4 $ SYM theory},
{JHEP {\bf 1512} (2015) 030}
[\href{http://arxiv.org/abs/1506.07551}
{\tt arXiv:1506.07551}]


\bibitem{Boels:2015yna}
R.~Boels, B.~A.~Kniehl and G.~Yang,
{\it Master integrals for the four-loop Sudakov form factor},
{Nucl.\ Phys.\ B {\bf 902} (2016) 387}
[\href{http://arxiv.org/abs/1508.03717}
{\tt arXiv:1508.03717}]

\bibitem{Huang:2016bmv}
R.~Huang, Q.~Jin and B.~Feng,
{\it Form Factor and Boundary Contribution of Amplitude},
{JHEP {\bf 1606} (2016) 072}
[\href{http://arxiv.org/abs/1601.06612}
{\tt arXiv:1601.06612}]

\bibitem{Maldacena:2010kp}
J.~Maldacena and A.~Zhiboedov,
{\it Form factors at strong coupling via a Y-system},
{JHEP {\bf 1011} (2010) 104}
[\href{http://arxiv.org/abs/1009.1139}
{\tt arXiv:1009.1139}]

\bibitem{Gao:2013dza}
Z.~Gao and G.~Yang,
{\it Y-system for form factors at strong coupling in $AdS_5$ and with multi-operator insertions in $AdS_3$},
{JHEP {\bf 1306} (2013) 105}
[\href{http://arxiv.org/abs/1303.2668}
{\tt arXiv:1303.2668}]

\bibitem{Wilhelm:2016izi}
M.~Wilhelm,
{\it Form factors and the dilatation operator in ${\cal N}\!=\!4$ super Yang-Mills theory and its deformations},
\href{http://arxiv.org/abs/1603.01145}
{\tt arXiv:1603.01145}

\bibitem{ArkaniHamed:2012nw}
N.~Arkani-Hamed, J.~L.~Bourjaily, F.~Cachazo, A.~B.~Goncharov, A.~Postnikov and J.~Trnka,
{\it Scattering Amplitudes and the Positive Grassmannian},
\href{http://arxiv.org/abs/1212.5605}
{\tt arXiv:1212.5605}

\bibitem{Dixon:2004za}
L.~J.~Dixon, E.~W.~N.~Glover and V.~V.~Khoze,
{\it MHV rules for Higgs plus multi-gluon amplitudes},
{JHEP {\bf 0412} (2004) 015}
[\href{http://arxiv.org/abs/hep-th/0411092}
{\tt hep-th/0411092}]


\bibitem{Eden:2011yp}
B.~Eden, P.~Heslop, G.~P.~Korchemsky and E.~Sokatchev,
{\it The super-correlator/super-amplitude duality:~Part I},
{Nucl.\ Phys.\ B {\bf 869} (2013) 329}
[\href{http://arxiv.org/abs/1103.3714}
{\tt arXiv:1103.3714}]

\bibitem{Eden:2011ku}
B.~Eden, P.~Heslop, G.~P.~Korchemsky and E.~Sokatchev,
{\it The super-correlator/super-amplitude duality:~Part II},
{Nucl.\ Phys.\ B {\bf 869} (2013) 378}
[\href{http://arxiv.org/abs/1103.4353}
{\tt arXiv:1103.4353}]

\bibitem{Dolan:2014ega}
L.~Dolan and P.~Goddard,
{\it The Polynomial Form of the Scattering Equations},
{JHEP {\bf 1407} (2014) 029}
[\href{http://arxiv.org/abs/1402.7374}
{\tt arXiv:1402.7374}]

\bibitem{Weinzierl:2014ava}
S.~Weinzierl,
{\it Fermions and the scattering equations},
{JHEP {\bf 1503} (2015) 141}
[\href{http://arxiv.org/abs/1412.5993}
{\tt arXiv:1412.5993}]

%
%

\bibitem{ArkaniHamed:2009dn}
N.~Arkani-Hamed, F.~Cachazo, C.~Cheung and J.~Kaplan,
{\it A Duality For The S Matrix},
{JHEP {\bf 1003} (2010) 020}
[\href{http://arxiv.org/abs/0907.5418}
{\tt arXiv:0907.5418}]

\bibitem{ArkaniHamed:2009si}
N.~Arkani-Hamed, F.~Cachazo, C.~Cheung and J.~Kaplan,
{\it The S-Matrix in Twistor Space},
{JHEP {\bf 1003} (2010) 110}
[\href{http://arxiv.org/abs/0903.2110}
{\tt arXiv:0903.2110}]

\bibitem{ArkaniHamed:2009dg}
N.~Arkani-Hamed, J.~Bourjaily, F.~Cachazo and J.~Trnka,
{\it Unification of Residues and Grassmannian Dualities},
{JHEP {\bf 1101} (2011) 049}
[\href{http://arxiv.org/abs/0912.4912}
{\tt arXiv:0912.4912}]

\bibitem{Koster:2016loo}
L.~Koster, V.~Mitev, M.~Staudacher and M.~Wilhelm,
{\it All Tree-Level MHV Form Factors in $\mathcal{N}\!=\!4$ SYM from Twistor Space},
{JHEP {\bf 1606} (2016) 162}
[\href{http://arxiv.org/abs/1604.00012}
{\tt arXiv:1604.00012}]

\bibitem{Koster:2016ebi}
L.~Koster, V.~Mitev, M.~Staudacher and M.~Wilhelm,
{\it Composite Operators in the Twistor Formulation of ${\cal N}\!=\!4$ Supersymmetric Yang-Mills Theory},
{Phys.\ Rev.\ Lett.\  {\bf 117} (2016) 011601}
[\href{http://arxiv.org/abs/1603.04471}
{\tt arXiv:1603.04471}]

\bibitem{Chicherin:2016soh}
D.~Chicherin and E.~Sokatchev,
{\it Demystifying the twistor construction of composite operators in ${\cal N}\!=\!4$ super-Yang-Mills theory},
\href{http://arxiv.org/abs/1603.08478}
{\tt arXiv:1603.08478}

\bibitem{Chicherin:2014uca}
D.~Chicherin, R.~Doobary, B.~Eden, P.~Heslop, G.~P.~Korchemsky, L.~Mason and E.~Sokatchev,
{\it Correlation functions of the chiral stress-tensor multiplet in ${\cal N}\!=\!4$ SYM},
{JHEP {\bf 1506} (2015) 198}
[\href{http://arxiv.org/abs/1412.8718}
{\tt arXiv:1412.8718}]

\bibitem{Brandhuber:2016xue}
A.~Brandhuber, E.~Hughes, R.~Panerai, B.~Spence and G.~Travaglini,
{\it The connected prescription for form factors in twistor space},
\href{http://arxiv.org/abs/1608.03277}
{\tt arXiv:1608.03277}


\end{thebibliography}
\end{document}